\documentclass[12pt]{iopart}
\usepackage{epsfig}

\begin{document}

\title{Formation of Quark Phases in compact stars and their connection to Gamma-Ray-Bursts}

\author{A. Drago$^a$, G. Pagliara$^b$, J. Schaffner-Bielich$^b$ }

\address{$^a$ Dipartimento di Fisica - Universit\`a di Ferrara and INFN Sez. di Ferrara,
44100 Ferrara, Italy}

\address{$^b$ Institut f\"{u}r Theoretische Physik, Goethe Universit\"{a}t,
  D-60438, Frankfurt am Main, Germany}

\begin{abstract}
We analyse the occurrence of quiescent times in the temporal 
structure of the Gamma-Ray-Bursts (GRBs) light
curves. We show that if a long quiescent time is present, it is
possible to divide the total duration of GRBs into three periods: the
pre-quiescence emission, the quiescent time and the post-quiescence
emission.  We then discuss a model of the GRBs inner engine based on
the formation of quark phases during the life of an hadronic
star. Within this model the pre-quiescence emission is interpreted as
due to the deconfinement of quark inside an hadronic star and the
formation of 2SC quark matter.  The
post-quiescence emission is due to the conversion of 2SC into
the Color-Flavor-Locking (CFL) phase.  The temporal delay between
these two processes is connected with the nucleation time of the CFL
phase in the 2SC phase and it can be associated with the observed
quiescent times in the GRBs light curves. The stability of CFL cores
in compact stars is also discussed.
\end{abstract}



\section{Introduction}

The time structure of GRBs is usually complex and
it often displays, during the phase of the prompt emission, 
several short pulses separated by time intervals
lasting from fractions of second to several tens of seconds. In some
cases very long period of vanishing signal, 
the so called {\it quiescent
times} (QTs) are present, which can have durations
comparable with the durations of the emission periods. 
Within the internal shock model QTs shorter than few tens of seconds can be explained by 
the modulation of a continuous shells emission
from the Inner Engine (IE) \cite{ramirez}.
When a long QT is present in the light curve, as we will
show, it is more plausible to assume that the inner engine
is dormant.
We discuss then the quark deconfinement model of the GRBs. 
In this model the interpretation of long QTs 
is given in terms of time intervals between 
readjustments of the structure of a compact star. We suggest that
the GRB emissions are due to first order phase transitions occurring
between different phases of the strong interacting matter and we associate
the periods of dormancy with the nucleation times needed to trigger
the phase transitions.

\section{Quiescent times in the GRBs prompt emission}

\begin{figure}
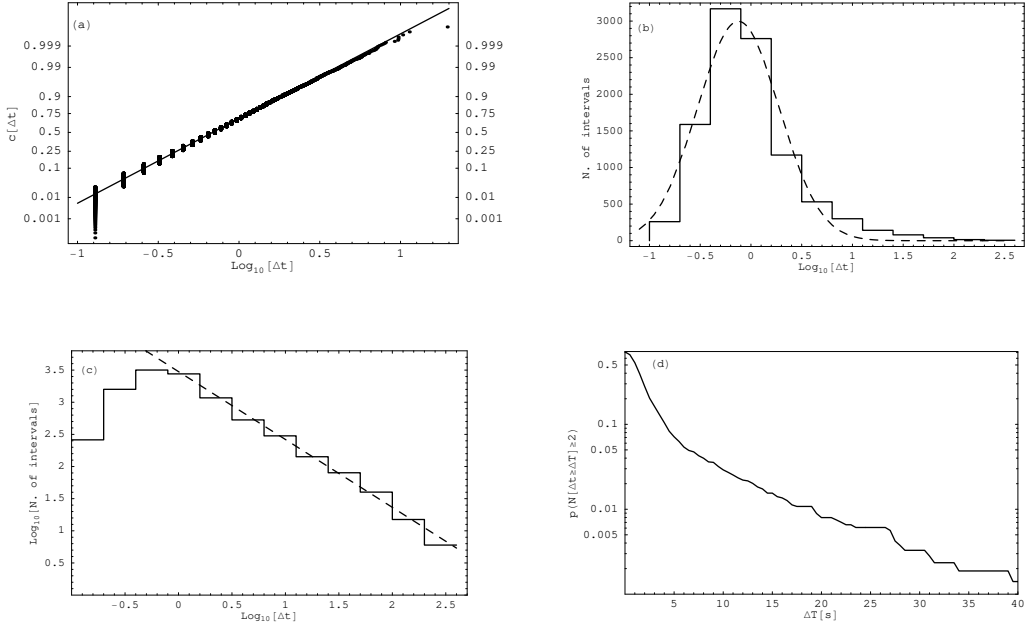

    \begin{centering}
\hbox{\hskip-0.cm \epsfig{file=f1a.epsi,height=3.6cm}\hskip 1.cm 
\epsfig{file=f1b.epsi,height=3.6cm}}
\vskip 1.cm    
\hbox{\hskip 0.3cm
\epsfig{file=f1c.epsi,height=3.6cm}
\hskip 1.2cm \epsfig{file=f1d.epsi,height=3.6cm}}
    \caption{{\bf  Analysis of time intervals between peaks} 
{\bf a} The cumulative distribution of time intervals $\Delta t$ which are not QTs
(black point), is compared with its best fit log-normal distribution (solid
line). {\bf b} Histogram
of the QTs and its log-normal fit (dashed line).
{\bf c} Histogram of QTs and power-low fit of its tail (dashed line).
{\bf d} Frequency of bursts containing at least two QTs longer than $\Delta T$.
\label{rit} }
   \end{centering}
   \end{figure}

A previous statistical analysis \cite{NP} has shown that there are
three time-scales in the GRB light curves: the shortest one is the
variability scale determining the pulses durations and the intervals
between pulses; the largest one describes the total duration of the
bursts and, finally, an intermediate time scale is associated with long
periods within the bursts having no activity, the QTs. 
The origin of these periods of quiescence is still unclear.
We have recently ~\cite{doppi} performed a new statistical analysis of the time intervals $\Delta
t$ between adjacent peaks in the light curve of GRBs using the algorithm introduced in Ref.~\cite{NP}.
We have applied this analysis to all the light curves of the BATSE
catalogue. In a first investigation we have merged all the bursts of the
catalogue into one sample from which we compute the cumulative
probability $c(\Delta t)$ of finding time intervals $\Delta t$ which are not
QTs i.e.  we compute the distribution of the time intervals within each active period.  
In Fig.~\ref{rit}a, we show that $c(\Delta t)$ is well described by a
log-normal distribution. In Fig.~\ref{rit}b, the histogram of QTs is
displayed together with a log-normal distribution. As already observed
by previous authors \cite{NP}, there is an evident deviation of the
data points respect to the log-normal distribution for time intervals
longer than a few seconds, indicating an excess of long $\Delta t$.
In Fig.~\ref{rit}c we show a power law fit of the tail of the QTs
distribution which displays a very good agreement with the data, as
already observed by \cite{quilligan}.  
Finally, in Fig.~\ref{rit}d we show a correlation function, indicating
the probability of finding at least 2 QTs longer than $\Delta T$ in a
same GRB.This probability rapidly decreases
and it essentially vanishes for $\Delta T > 40$ s.

We can now define a subsample of the BATSE catalogue composed of all
the bursts having a QT longer than $40$ s and study its properties.
From the result of Fig.~\ref{rit}d, the bursts of the subsample contain only one
long QT and it is therefore possible to divide each burst into a pre-quiescence emission
(PreQE) and a post-quiescence emission (PostQE) of which we will compare the temporal and spectral structure.


\begin{figure}
\begin{center}
\includegraphics[scale=0.5]{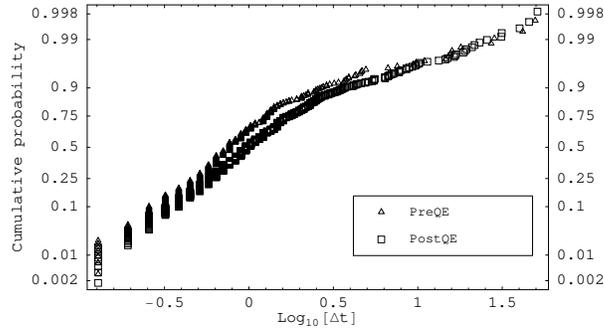}
\end{center}
\caption{{\bf Analysis of time intervals between peaks within the two emission periods} 
The cumulative distributions of $\Delta t$ are shown for the two emission episodes. 
\label{twoemission} }
\end{figure}

In Fig.~\ref{twoemission} we display the cumulative distributions
$c_1(\Delta t)$ and $c_2(\Delta t)$ within each of the two emission
periods. The two distributions are very similar as also confirmed by the $\chi^2$-test.
Let us remind that within the
internal-external-shocks model \cite{piran2,meszaros}, external shocks produce
emissions lacking the short time scale variability produced by
internal shocks \cite{saripiran}. The result of Fig.~\ref{twoemission}
rules out a scenario in which PostQE is dominated by external shocks
and PreQE by internal shocks.  This in turn excludes the possibility
of associating the QTs with the time needed for the jet to reach and
interact with the interstellar medium.

The statistical analysis of the durations $D1$ and $D2$
of the two emission periods (see Ref.~\cite{doppi}) show that the
two data sets are well fitted by two log-normal distributions.  The
two distributions have different mean values ($D1_{ave}\sim 21s$,
$D2_{ave}\sim 41s$) 
and almost
identical standard deviations ($\sigma_1=36s$, $\sigma_2=33s $).
To estimate the emitted energy during PreQE and PostQE we have
analysed the hardness ratios and the  power emitted in each emission. Both quantities
are on average the same.
Let us now discuss the implications of this analysis on the origin of QTs.
As observed by \cite{ramirez},
within the internal shocks model it is possible to explain the QTs
either as a turn-off of the IE or as a modulation of a
continuous relativistic wind emitted by the IE
(Wind Modulation Model WMM). Both
hypothesis are consistent with the result of
Fig.~\ref{twoemission}.
The main difference between the WMM and the dormant engine scenario
is that in the WMM the inner engine has to provide a constant
power during the whole duration of the burst. 
In our subsample, we have several bursts whose total duration
(including the QT) approaches 300 s. 
These durations have to be corrected taking into account the average
redshift of the BATSE catalogue, $z_{ave} \sim 2 $, but
even after this renormalization, durations of a hundred seconds or more are not
too rare. 
This time scale has to be compared with the typical duration of the
emission period of the inner engine, as estimated in various models.
For instance, in 
all numerical investigations of the collapsar model \cite{woosley}
the IE remains switched-on
during some 20s.  
Also in the quark deconfinement model which will
be discussed in the next section
the inner engine remains active during periods of the order of
a few ten seconds corresponding to the cooling time of the
compact stellar object. We conclude therefore that the energy requirement within the WMM
scenario is too large with respect to the results of the theoretical models.
It is instead more plausible to assume that during a QT the inner engine
switches off, sometimes for very long periods, and then 
restart producing a PostQE very similar to the PreQE.

\section{Phase transitions between QCD phases and CFL core in compact stars}

Let us now discuss how to generate dormancy periods in the prompt
emission within the quark deconfinement model.
In this model the energy source powering the GRB is
the transition from a star containing only hadrons to a star composed, at least in part, of 
deconfined quarks \cite{apj}.
In the first calculations within the quark deconfinement model, the
equation of state (EoS) of quarks 
was computed using the MIT bag model.
Actually, in the last years the 
possibility of forming
a diquark condensate at the center of a compact star has been widely
discussed in the literature \cite{raj}. It was shown that the formation of a color
superconducting quark core can increase the energy released by a
significant amount \cite{noiprd}. In particular, many calculations indicate that Color-Flavor-Locked (CFL) quark matter is the
most stable configuration at large density whereas at intermediate
density the two flavor color superconducting 2SC phase, or the normal
quark (NQ) phase depending on the parameters, are favoured \cite{ruster1,david0}.
The transition
from the 2SC matter to CFL matter can take place as a first order phase transition if the
leptonic content of the newly formed normal quark matter phase is not too small
\cite{ruster2}. It is therefore tempting to associate the
PreQE with the transition from the hadronic to the 2SC phase and the
PostQE with the formation of the CFL phase \cite{noi}.  In
this scenario the two dimensional scales regulating the durations of
PreQE and PostQE are the energies released in the two transitions.
Concerning the conversion process between different QCD phases, it turns out 
that the conversion always takes place as a strong deflagration and
never as a detonation \cite{parenti}. This is important because
in the case of a detonation the region in which the 
electron-photon plasma forms (e.g. via neutrino-antineutrino annihilation near the
surface of the compact star) would be contaminated by the baryonic load 
and it would be impossible to accelerate the plasma up to the enormous Lorentz factors
needed to explain the GRBs.

\begin{figure}
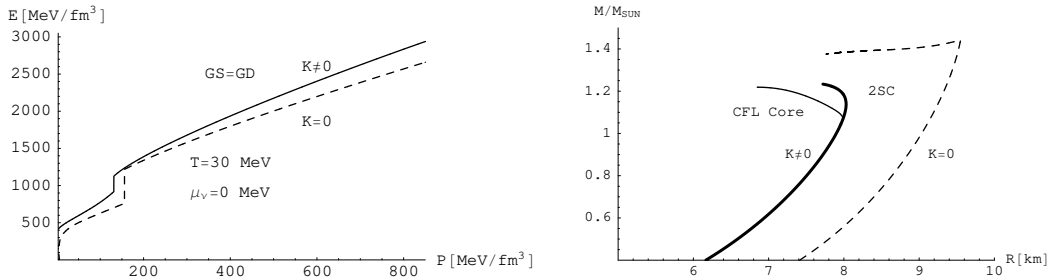

    \begin{centering}
\hbox{\hskip-0.cm \epsfig{file=f3a.epsi,height=3.6cm}\hskip 0.7cm 
\epsfig{file=f3b.epsi,height=3.6cm}}
    \caption{{\bf Equation of state and mass-radius relation} 
Left panel:The energy density as a function of the pressure is shown for the two
cases with and without the six-fermions interaction term
($K\neq0$, $K=0$) for T=30 MeV and  $\mu_{\nu}=0$. The diquark coupling constant is
chosen to be equal the the quark-antiquark coupling constant (see \cite{ruster1,ruster2}).
Right panel: mass-radius relation for quark cores. The thin and thick lines correspond
to the EoS with $K\neq0$ with a transition from 2SC to the CFL
phase and for only the 2SC phase respectively, the dashed line stands for the case
$K=0$. When the CFL transition is reached in the first case a
stable CFL core is possible; instead in the second case this leads to
an instability of the star.
\label{mass} }
   \end{centering}
   \end{figure}

We want to discuss now the possible formation of a CFL phase core inside compact stars.
While it has been shown in MIT-inspired quark models that the CFL phase can appear 
in hybrid stars or purely quark stars \cite{alford,noi}, 
up to now, this is not the case within NJL-inspired models:
in hybrid stars the appearance of a CFL core renders the star unstable \cite{buballa,david1}.
Actually, in very recent study of the quark core of proto-neutron stars with the NJL model 
it has been show that in a tiny window of baryonic masses and for vanishing temperature and neutrino 
chemical potential a CFL core can form \cite{david2}. Nevertheless the conclusion of that 
work is that by taking into account the influence of the hadronic crust, the hybrid stars
become unstable again when the central density reaches the onset of the CFL phase transition.
As in Ref.~\cite{david2}, we want to discuss the possible formation of
a CFL core in a proto-neutron star limiting the discussion to the EoS
of the quark phase. A complete study including also
the hadronic phase and the Gibbs construction, as made in Ref.~\cite{jurgen}, 
for the phase transition is still in progress. To compute the EoS
of quark matter we use the NJL-like model of
Ref.\cite{ruster1,ruster2}. The only difference with respect to the model
discussed in Ref.~\cite{david2} is the inclusion of the six-fermion interaction term in
the Lagrangian which simulates the breaking of axial
symmetry. As already observed in Ref.~\cite{buballa}, this term is
responsible for the flavor-mixing of quarks which in turn leads to a
different behaviour of the strange quark mass $m_s$ as a function of the chemical potential : in the absence of
the six-fermions interaction (i.e. choosing $K=0$), $m_s$ has the
same value both in the chiral broken phase and in the intermediate 2SC (NQ)
phase and then rapidly decreases at
the onset of the CFL phase \cite{david1}. In the case of $K\neq0$,  
$m_s$ jumps to lower values already in the 2SC
(NQ) phase and then decreases
again in the CFL phase \cite{ruster1}. This different behaviour has
noticeable effects on the EoS. As already observed by
Buballa \cite{buballa}, for $K\neq0$ the phase transition from the 2SC to
the CFL phase occurs at lower density. Moreover, the
gap of the energy density at the onset of the 2SC-CFL transition is considerably
lower when the six-fermion term is included as we can observe in the
left panel of Fig.~\ref{mass} 
where the EoS of quark matter is shown for a temperature of 30 MeV and vanishing 
neutrino chemical potential $\mu_{\nu}$.  
This feature has interesting effects on the corresponding mass-radius relation of compact
stars. In the right panel of Fig.~\ref{mass}, we show the mass-radius relations using the
EoS discussed before for the cases $K\neq0$ and $K=0$, with $T=30$ MeV, $\mu_{\nu}=0$. 
CFL cores do exist for $K\neq0$  in a sizeable window of
masses. This is a consequence of the weaker phase transition
between the 2SC and the CFL phases when the six-fermion interaction is included. 
It also interesting to observe that the CFL core can form already at
large temperatures while in the case  $K=0$ it can form only at
vanishing temperatures. This could have important consequences for the
evolution of the proto-neutron star. Computing
the differences of gravitational masses between
a 2SC and a CFL star having the same baryonic mass,  values
of few per cent of solar mass can be obtained which correspond to energies
released in the conversion of the order of $10^{52}$ ergs. In our model this process powers the PostQE
emission.

\section{Conclusions}
We have provided hints for the interpretation of long QTs in GRB
as due to dormancy periods of the inner engine. Before and after a long quiescent time
the temporal microstructures of the emissions are similar. Interestingly, the average duration 
of PostQE is longer than the average duration of PreQE
indicating that probably a larger amount of energy is released by the inner engine in the PostQE.
Within the quark deconfinement model is possible to associate the PreQE with the 
deconfinement phase transition in which the 2SC phase is formed and the PostQE with the formation
of the CFL phase. It has also been shown, considering only the quark EoS,
that stable CFL cores can be formed in compact stars. On this point a more
refined calculation is needed matching the quark EoS at
large density with the hadronic EoS
at low density.

\vspace{0.15cm}
{\bf Acknowledgements}:\\
We thank S. R\"{u}ster and D. Rischke for providing us the code to compute the equation of state of quark matter. 
G.P. thanks INFN for financial support.

\end{document}